\pdfoutput=1
\documentclass[
  reprint,
  superscriptaddress,
 amsmath,amssymb,
 aps,
 prl
]{revtex4-1}

\usepackage{graphicx}
\usepackage{dcolumn}
\usepackage{bm}
\usepackage{color}

\begin{document}

\title{
  Optically tunable spin Hall effect
  in periodically driven monolayer transition metal dichalcogenides
}

\author{Naoya Arakawa}
\email{arakawa@phys.chuo-u.ac.jp}
\affiliation{The Institute of Science and Engineering,
  Chuo University, Bunkyo, Tokyo, 112-8551, Japan}
\author{Kenji Yonemitsu}
\affiliation{The Institute of Science and Engineering,
  Chuo University, Bunkyo, Tokyo, 112-8551, Japan}
\affiliation{Department of Physics,
  Chuo University, Bunkyo, Tokyo 112-8551, Japan}


\begin{abstract}
  We show that
  the driving field of 
  circularly polarized light (CPL)
  can be used to enhance and reverse 
  the spin current generated in the spin Hall effect
  for some transition-metal dichalcogenides. 
  This is demonstrated
  by analyzing the time-averaged spin Hall conductivities
  in the nonequilibrium steady states of monolayers
  WS$_{2}$, MoS$_{2}$, MoTe$_{2}$, and WTe$_{2}$ driven by CPL
  with the Floquet linear-response theory.
  We argue that 
  the enhancement and reversals of the spin current
  come from a combination of the non-Rashba effect of broken inversion symmetry 
  and the nonperturbative effect of CPL
  beyond the dynamical localization.
  This work allows optical control of
  the magnitude and direction of the pure spin current.
\end{abstract}
\maketitle

\begin{figure}
  \includegraphics[width=86mm]{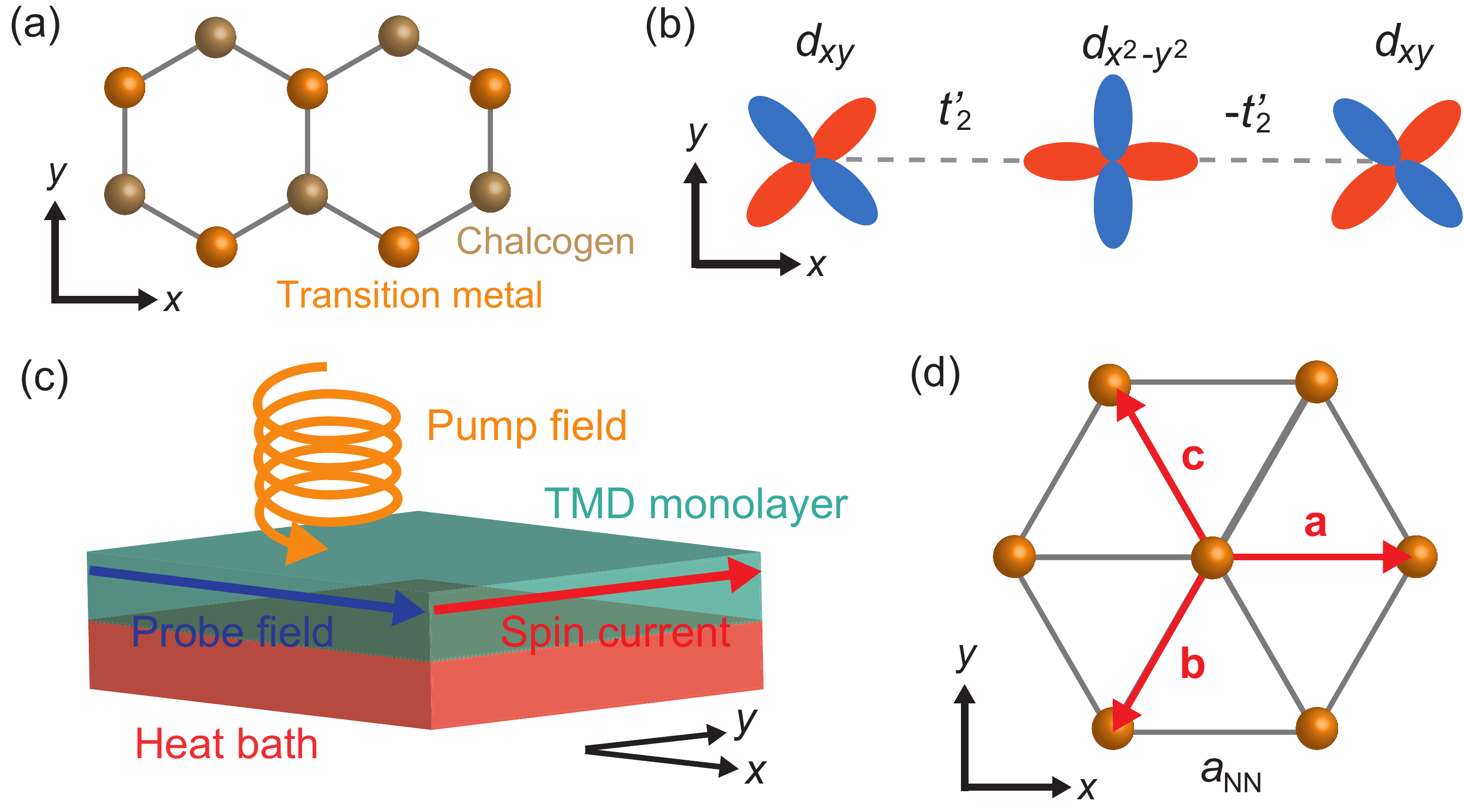}
  \caption{\label{fig1}
    (a) The honeycomb lattice formed by transition metal and chalcogen ions.
    (b) The parity-odd interorbital hopping integrals
    between the $d_{xy}$ and $d_{x^{2}-y^{2}}$ orbitals. 
    The color difference represents
    the difference in the signs of the wave functions.
    (c) The set-up for the SHE of the monolayer TMDs
    driven by CPL.
    (d) The triangular lattice formed by transition metal ions
    with three nearest-neighbor vectors.
  }
\end{figure}

Transition metal dichalcogenides (TMDs) have provided
opportunities for many attractive phenomena.
For example,
monolayer TMDs can be used to realize various valleytronics phenomena.
In some monolayer TMDs~\cite{TMD1,TMD2,TMD3},
the monolayer consists of the upper and lower planes of chalcogen ions
and the middle plane of transition metal ions
with a trigonal prismatic arrangement.
Because of this structure,
the monolayer TMDs can be approximately regarded as two-dimensional systems
on a honeycomb lattice [Fig. \ref{fig1}(a)]
with broken inversion symmetry~\cite{TMD-ISB1,TMD-ISB2}.
Since the valley degeneracy can be lifted
with broken inversion and time-reversal symmetries,
valley polarization~\cite{TMD-ISB1} and
a valley-selective Hall effect~\cite{valleyHall}
can be achieved by applying weak, resonant circularly polarized light (CPL).
Then,
the monolayer TMDs with another crystal structure
are suitable for the quantum spin-Hall insulator
(i.e., the topological insulator)~\cite{TopoIns-TMD1}.
This was
experimentally confirmed~\cite{TopoIns-TMD2}. 
Moreover, some combinations of TMDs form
the moir\'{e} systems~\cite{Moire-PRL,Moire-Nature1,Moire-Nature2,Moire-Nature3},
which have been intensively studied in recent years. 

Although some monolayer TMDs have the advantage of
realizing a pure spin current,
this advantage has not been fully taken yet.
Their broken inversion symmetry is characterized by 
parity-odd interorbital hopping integrals~\cite{TMD-TB1,TMD-TB2},
the signs of which change under an inversion operation [Fig. \ref{fig1}(b)],
whereas  
the Rashba spin-orbit coupling (SOC)~\cite{Rashba} is absent
due to the mirror symmetry of the $xy$ plane.
Therefore, the broken inversion symmetry can be described by the non-Rashba effect. 
Then,
the electronic states near the Fermi level 
can be well described by the $d_{3z^{2}-r^{2}}$, $d_{xy}$, and $d_{x^{2}-y^{2}}$ orbitals of
transition metal ions~\cite{TMD-1stPrin1,TMD-1stPrin2,TMD-1stPrin3},
which have the Ising-type SOC.
Because of these properties, 
the $z$ component of the spin angular momentum is conserved,
and, as a result,
the spin current is well defined.
This contrasts with the ill-defined spin current
in the presence of the Rashba SOC.
It is also in contrast to 
the coupled spin and orbital angular momenta 
in transition metals such as Pt~\cite{Kon-SHE-Pt}.
The well-defined spin current is advantageous to realize
a pure spin current.
However,
the monolayer TMDs at zero temperature
possess the negligibly small spin Hall effect (SHE)~\cite{TMD-SHE-0K},
in which a spin current is generated perpendicular to
an applied electric field~\cite{SHE-JETP,SHE-Hirsch}.

In this Letter, we show that
strong CPL can be used to enhance and reverse
the spin current generated in the SHE [Fig. \ref{fig1}(c)] for some monolayer TMDs.
Using 
the Floquet linear-response theory~\cite{Eckstein,Tsuji,Mikami,NA-FloquetSHE,NA-BCPL,NA-mirror}, 
we study the time-averaged spin Hall conductivity $\sigma_{yx}^{\textrm{S}}$ 
in the nonequilibrium steady states of monolayers WS$_{2}$, MoS$_{2}$, MoTe$_{2}$, and WTe$_{2}$
driven by CPL.
We show that
the magnitude and sign of $\sigma_{yx}^{\textrm{S}}$ can be changed
with increasing the magnitude of the CPL field.
The enhancement and sign changes disappear
if the parity-odd interorbital hopping integrals are zero. 
They are attributed to
a combination of the non-Rashba effect due to broken inversion symmetry and 
the nonperturbative effect of CPL beyond the dynamical localization.
We propose that
the periodically driven monolayer TMDs offer the optically tunable SHE.

\begin{figure*}
  \includegraphics[width=180mm]{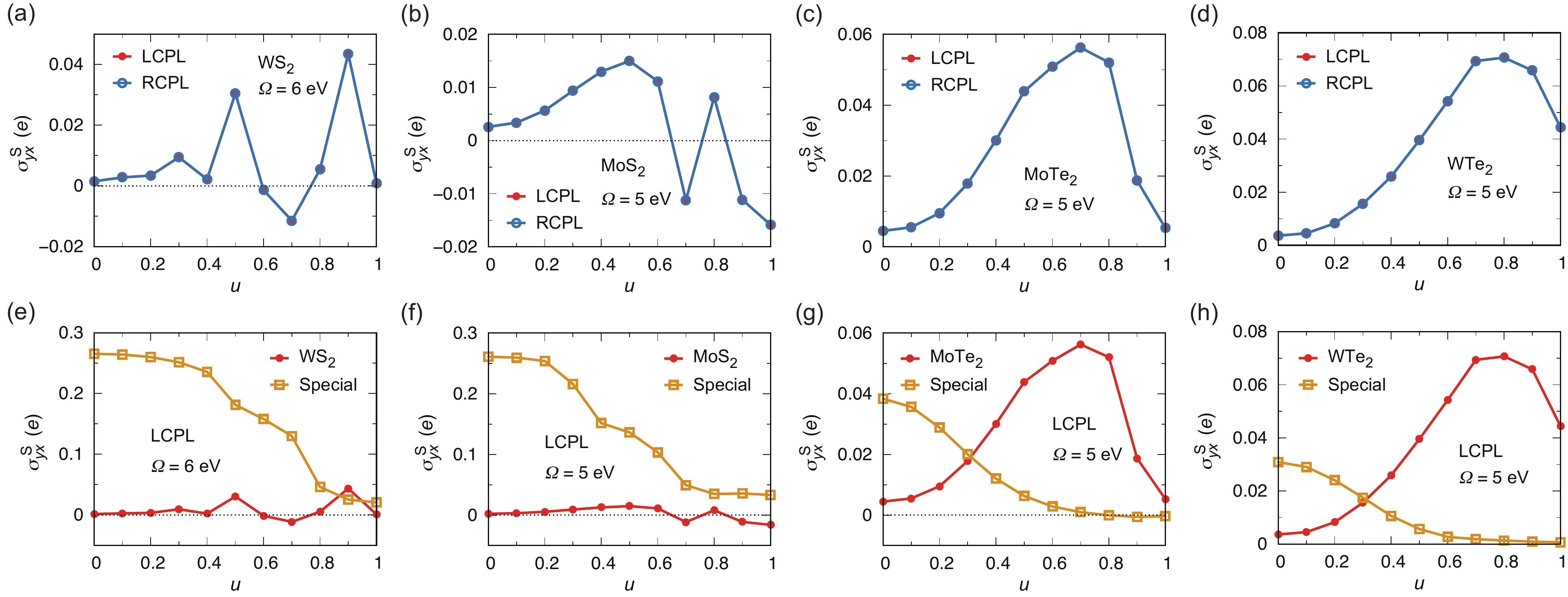}
  \caption{\label{fig2}
    $\sigma_{yx}^{\textrm{S}}$
    obtained in the Floquet linear-response theory
    for monolayers (a) WS$_{2}$, (b) MoS$_{2}$, (c) MoTe$_{2}$, and (d) WTe$_{2}$ driven
    by LCPL or RCPL.
    Each value of the light frequency $\Omega$ is larger than the bandwidth in each nondriven case.
    $\sigma_{yx}^{\textrm{S}}$
    obtained in the Floquet linear-response theory
    for the tight-binding models and special models
    of monolayers (e) WS$_{2}$, (f) MoS$_{2}$, (g) MoTe$_{2}$, and (h) WTe$_{2}$ driven
    by LCPL.
    In the numerical calculations,
      we set $n_{\textrm{max}}=2$, $T_{\textrm{b}}=0.02$ eV, and $\Gamma=0.01$ eV
      (see Supplemental Note 3 of Supplemental Material~\cite{SM}).
  }
\end{figure*}

\textit{Model.{--}}The periodically driven monolayer TMDs are described
by
$H=H_{\textrm{s}}(t)+H_{\textrm{b}}+H_{\textrm{sb}}$,
where $H_{\textrm{s}}(t)$ is the Hamiltonian for the monolayer TMDs driven by CPL,
$H_{\textrm{b}}$ is the Hamiltonian
for the B\"{u}ttiker-type heat bath~\cite{HeatBath1,HeatBath2} at temperature $T_{\textrm{b}}$,
and $H_{\textrm{sb}}$ is the system-bath coupling Hamiltonian.
We take a three-orbital tight-binding model~\cite{TMD-TB1} with the Peierls phase factor
due to the CPL field
$\mathbf{A}_{\textrm{pump}}(t)=(A_{0}\cos\Omega t\ A_{0}\sin(\Omega t+\delta))^{T}$
($\delta=0$ or $\pi$) as $H_{\textrm{s}}(t)$:
\begin{align}
  H_{\textrm{s}}(t)=
  \sum_{i,j}\sum_{a,b=1}^{3}\sum_{\sigma=\uparrow,\downarrow}
      [t^{ij}_{ab}(t)+\delta_{i,j}(\epsilon_{a}\delta_{a,b}+\xi_{ab}^{\sigma})]
      c_{ia\sigma}^{\dagger}c_{jb\sigma},\label{eq:Hs}
\end{align}
where 
$t^{ij}_{ab}(t)=t_{ab}^{ij}e^{-ie\mathbf{A}_{\textrm{pump}}(t)\cdot(\mathbf{R}_{i}-\mathbf{R}_{j})}$
is the hopping integral between transition metal ions
on a triangular lattice [Fig. \ref{fig1}(d)]
with the Peierls phase factor,
$\epsilon_{a}$ is the onsite energy,
$\xi_{ab}^{\sigma}=i\xi(\delta_{\sigma,\uparrow}-\delta_{\sigma,\downarrow})(\delta_{a,2}\delta_{b,3}-\delta_{a,3}\delta_{b,2})$ is the Ising-type SOC,
and $a=1$, $2$, or $3$ represent the $d_{3z^{2}-r^{2}}$, $d_{xy}$, and $d_{x^{2}-y^{2}}$ orbitals,
respectively.
Hereafter,
we set $\hbar=k_{\textrm{B}}=c=a_{\textrm{NN}}=1$,
where $a_{\textrm{NN}}$ represents the length between nearest neighbor sites. 
We have considered $H_{\textrm{b}}$ and $H_{\textrm{sb}}$
because the weak coupling to the bath can be used to achieve
a nonequilibrium steady state under the heating
due to $\mathbf{A}_{\textrm{pump}}(t)$~\cite{Tsuji,Mikami,NA-FloquetSHE,NA-BCPL,NA-mirror}.

As specific systems,
we consider monolayers WS$_{2}$, MoS$_{2}$, MoTe$_{2}$, and WTe$_{2}$.
To describe their electronic states,
we express $t_{ab}^{ij}$'s
in terms of six nearest-neighbor hopping integrals~\cite{TMD-TB1}, 
four parity-even ones $t_{1}$, $t_{2}$, $t_{3}$, and $t_{4}$
and two parity-odd ones $t_{1}^{\prime}$ and $t_{2}^{\prime}$,
and choose them, $\epsilon_{a}$'s, and $\xi$
in a similar way to that of Ref. \onlinecite{TMD-TB1}.
(For more details, see Supplemental Note 1 of
Supplemental Material~\cite{SM}.)

\textit{Optically tunable SHE.{--}}The SHE
for the periodically driven monolayer TMDs can be studied 
in the Floquet linear-response theory.
In this theory~\cite{Eckstein,Tsuji,Mikami,NA-FloquetSHE},
the pump and probe fields $\mathbf{A}_{\textrm{pump}}(t)$ and $\mathbf{A}_{\textrm{prob}}(t)$
are considered,
and the effects of $\mathbf{A}_{\textrm{pump}}(t)$ and $\mathbf{A}_{\textrm{prob}}(t)$
are treated in the Floquet theory~\cite{Floquet1,Floquet2}
and linear-response theory~\cite{Kubo}, respectively.
$\mathbf{A}_{\textrm{pump}}(t)$ is chosen to be the CPL field,
and $\mathbf{A}_{\textrm{prob}}(t)$ is applied along the $x$ axis 
to generate the spin current along the $y$ axis [Fig. \ref{fig1}(c)].
This spin current observed in the nonequilibrium steady states 
can be characterized by
the time-averaged spin Hall conductivity
$\sigma_{yx}^{\textrm{S}}$~\cite{NA-FloquetSHE}.
(For more details, see Supplemental Note 2 of Supplemental Material~\cite{SM}.)

Figures \ref{fig2}(a){--}\ref{fig2}(d)
show the dependences of $\sigma_{yx}^{\textrm{S}}$ numerically calculated in
the Floquet linear-response theory
on a dimensionless quantity $u=eA_{0}$ 
for monolayers WS$_{2}$, MoS$_{2}$, MoTe$_{2}$, and WTe$_{2}$ driven by
left- or right-handed CPL (LCPL or RCPL).
(For details of the numerical calculations, see
Supplemental Note 3 of Supplemental Material~\cite{SM}.)
First, $\sigma_{yx}^{\textrm{S}}$ is independent of the helicity of light.
This is the same as that obtained
in an inversion-symmetric multiorbital system~\cite{NA-FloquetSHE} driven by CPL,
and it is due to the time-reversal symmetry of the spin current~\cite{NA-FloquetSHE}.
Then,
$\sigma_{yx}^{\textrm{S}}$ shows the non-monotonic $u$ dependences:
in all the cases, 
$\sigma_{yx}^{\textrm{S}}$ can be enhanced;
only in the cases of MoS$_{2}$ and WS$_{2}$ can 
the sign of $\sigma_{yx}^{\textrm{S}}$ be reversed.
The similar changes in magnitude and sign are obtained
even if $\Omega$ is smaller
(see Supplemental Note 4 of Supplemental Material~\cite{SM}).
We should note that 
the sign changes in $\sigma_{yx}^{\textrm{S}}$ can be achieved
in the cases of MoTe$_{2}$ and WTe$_{2}$
for $\Omega=3$ eV (see Supplemental Note 4 of
Supplemental Material~\cite{SM}).
These results suggest that
the spin current generated in the SHE can be enhanced and reversed
by tuning the CPL field. 
Such nonmonotonic $u$ dependences contrast
the monotonically decreasing $u$ dependence obtained
in the inversion-symmetric multiorbital system driven by CPL~\cite{NA-FloquetSHE}. 
Since that monotonic $u$ dependence can be understood as
the dynamical localization (i.e., the reduction in the kinetic energy) due to the CPL field,
the enhancement of $\sigma_{yx}^{\textrm{S}}$ and its sign changes
can be interpreted as 
the nonperturbative effects of CPL beyond the dynamical localization.
Note that the finite small $\sigma_{yx}^{\textrm{S}}$ in MoS$_{2}$ at $u=0$ does not contradict
the zero-temperature result~\cite{TMD-SHE-0K}
because
the finite-temperature effects, such as the broadening of the distribution function,
are taken into account in our calculations. 

\textit{Importance of the non-Rashba effect{--}}To
understand the effect of the parity-odd interorbital hopping integrals
on $\sigma_{yx}^{\textrm{S}}$,
we perform similar analyses in special models having inversion symmetry,
in which $t_{1}^{\prime}=t_{2}^{\prime}=0$,
i.e., the non-Rashba effect of broken inversion symmetry is absent. 
(For details of the special models,
see Supplemental Note 5 of Supplemental Material~\cite{SM}.)
Figures \ref{fig2}(e){--}\ref{fig2}(h) compare the $u$ dependences of $\sigma_{yx}^{\textrm{S}}$
numerically calculated in the Floquet linear-response theory
for the original and special models of monolayers WS$_{2}$, MoS$_{2}$, MoTe$_{2}$, and WTe$_{2}$
driven by LCPL.
In the special models,
$\sigma_{yx}^{\textrm{S}}$ monotonically decreases with increasing $u$.
This suggests that 
the parity-odd interorbital hopping integrals are vital
for achieving the nonmonotonic $u$ dependence of $\sigma_{yx}^{\textrm{S}}$.
(Note that
the large $\sigma_{yx}^{\textrm{S}}$'s at $u=0$ in the special models
are due to the tiny gaps at the K and K' points.)

The importance of the parity-odd interorbital hopping integrals
is supported by the analyses using the high-frequency expansion.
If the light frequency is off-resonant,
the periodically driven system
can be approximately described by an effective time-independent Hamiltonian
obtained in the high-frequency expansion~\cite{Highw1,Highw2,Mikami},
$H_{\textrm{eff}}= H_{0}+\Delta H$,
where
$H_{n}=\int_{0}^{T_{\textrm{p}}}\frac{dt}{T_{\textrm{p}}}e^{in\Omega t}H_{\textrm{s}}(t)$
and $\Delta H\approx \frac{[H_{-1},H_{1}]}{\Omega}$.
Note that the dynamical localization is described by $H_{0}$,
whereas the nonperturbative effect beyond it is described by $\Delta H$. 
For the monolayer TMDs driven by CPL,
$\Delta H$ consists of 
the light-induced hopping integrals between
nearest and next nearest neighbor sites,
$\Delta H=\Delta H_{\textrm{NN}}+\Delta H_{\textrm{NNN}}$,
and each term 
can be categorized as either a parity-even or a parity-odd term
(see Supplemental Note 6 of Supplemental Material~\cite{SM}).
Then,
the analyses of 
the group velocities induced by $\Delta H_{\textrm{NN}}$ or $\Delta H_{\textrm{NNN}}$
imply that 
the light-induced parity-odd hopping integrals,
which are finite with the parity-odd interorbital hopping integrals,
lead to the correction terms to the spin Hall conductivity
(for details, see Supplemental Note 7 of Supplemental Material~\cite{SM}).
This is consistent with the vital role of 
the parity-odd interorbital hopping integrals
and the nonperturbative effect beyond the dynamical localization.

The same conclusion is reached by analyzing
  the group velocities calculated in the Floquet theory.
  As we show in Supplemental Note 8 of Supplemental Material~\cite{SM},
  the light-induced leading-order corrections to $\sigma_{yx}^{\textrm{S}}$ 
  are proportional to
  the products of the parity-odd and parity-even hopping integrals,
  and they cause the new contributions to $\sigma_{yx}^{\textrm{S}}$
  from the momenta different from those at which
  the nondriven terms give the contributions.
  These results also support the important role of the non-Rashba effect. 

\begin{figure}
  \includegraphics[width=86mm]{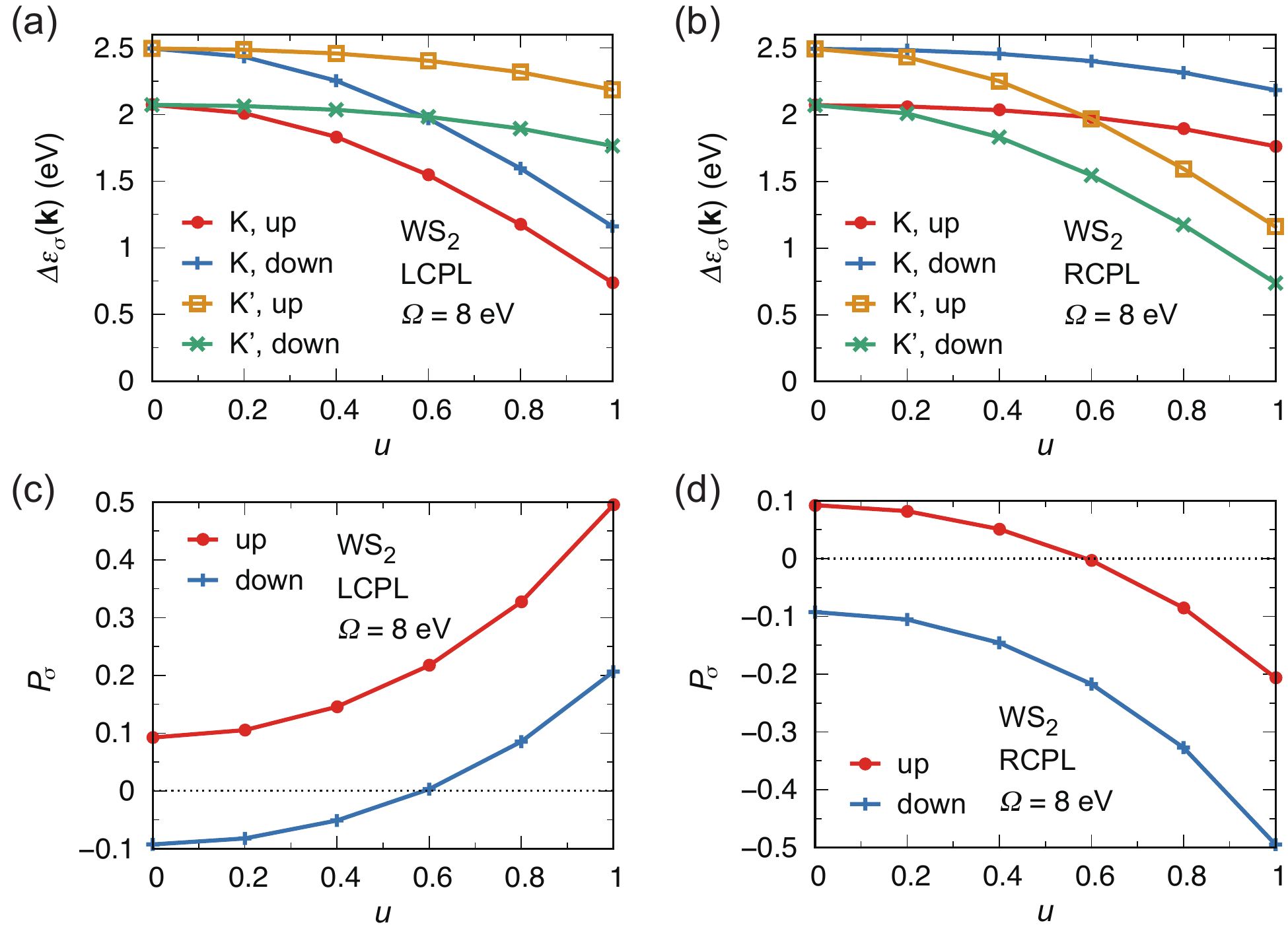}
  \caption{\label{fig3}
    $\Delta\epsilon_{\uparrow}(\mathbf{k}_{\textrm{K}})$,
    $\Delta\epsilon_{\downarrow}(\mathbf{k}_{\textrm{K}})$,
    $\Delta\epsilon_{\uparrow}(\mathbf{k}_{\textrm{K}^{\prime}})$,
    and $\Delta\epsilon_{\downarrow}(\mathbf{k}_{\textrm{K}^{\prime}})$ obtained in
    the high-frequency expansion for monolayer WS$_{2}$ driven by
    (a) LCPL or (b) RCPL.
    Here $\mathbf{k}_{\textrm{K}}=(\frac{4\pi}{3}\ 0)^{T}$,
    and $\mathbf{k}_{\textrm{K}^{\prime}}=(\frac{8\pi}{3}\ 0)^{T}$, which is equivalent to
    $(-\frac{4\pi}{3}\ 0)^{T}$. 
    $P_{\uparrow}$ and $P_{\downarrow}$
    obtained in the high-frequency expansion for WS$_{2}$
    driven by (c) LCPL or (d) RCPL.
  }
\end{figure}

\textit{Discussion.{--}}
First, we
discuss the valley degeneracy for our periodically driven monolayer TMDs.
Figure \ref{fig3}(a) or \ref{fig3}(b) shows the $u$ dependences of
$\Delta\epsilon_{\sigma}(\mathbf{k}_{\textrm{K}})$
and $\Delta\epsilon_{\sigma}(\mathbf{k}_{\textrm{K}^{\prime}})$
calculated in the high-frequency expansion
for monolayer WS$_{2}$ driven by LCPL or RCPL, respectively, 
at $\Omega=8$ eV.
Here $\Delta\epsilon_{\sigma}(\mathbf{k}_{\textrm{K}})$
or $\Delta\epsilon_{\sigma}(\mathbf{k}_{\textrm{K}^{\prime}})$ is
the energy difference
between the bottom of the lowest unoccupied band and
the top of the highest occupied band 
for spin $\sigma$ at the K or K$^{\prime}$ points, respectively.  
At $u=0$,
$\Delta\epsilon_{\uparrow}(\mathbf{k}_{\textrm{K}})=\Delta\epsilon_{\downarrow}(\mathbf{k}_{\textrm{K}^{\prime}})$
and
$\Delta\epsilon_{\downarrow}(\mathbf{k}_{\textrm{K}})=\Delta\epsilon_{\uparrow}(\mathbf{k}_{\textrm{K}^{\prime}})$.
Meanwhile, for $u\neq 0$ with LCPL,
$\Delta\epsilon_{\uparrow}(\mathbf{k}_{\textrm{K}})<\Delta\epsilon_{\downarrow}(\mathbf{k}_{\textrm{K}^{\prime}})$
and
$\Delta\epsilon_{\downarrow}(\mathbf{k}_{\textrm{K}})<\Delta\epsilon_{\uparrow}(\mathbf{k}_{\textrm{K}^{\prime}})$;
with RCPL,
these inequality signs are reversed.
The same properties hold for the other monolayer TMDs driven by CPL
(see Supplemental Note 9 of Supplemental Material~\cite{SM}). 
These results indicate that
the valley degeneracy is lifted if and only if
time-reversal symmetry, as well as inversion symmetry, is broken.
This is the same as that obtained 
in graphene driven by bicircularly polarized light~\cite{NA-BCPL}.

We also discuss the valley polarization. 
Figure \ref{fig3}(c) or \ref{fig3}(d) shows the $u$ dependence of $P_{\sigma}$, 
the valley polarization for spin $\sigma$, 
calculated in the high-frequency expansion for WS$_{2}$ driven by
LCPL or RCPL, respectively, at $\Omega=8$ eV,
where 
$P_{\sigma}=\frac{-\Delta\epsilon_{\sigma}(\mathbf{k}_{\textrm{K}})+\Delta\epsilon_{\sigma}(\mathbf{k}_{\textrm{K}^{\prime}})}{\Delta\epsilon_{\sigma}(\mathbf{k}_{\textrm{K}})+\Delta\epsilon_{\sigma}(\mathbf{k}_{\textrm{K}^{\prime}})}$. 
Note that $P_{\sigma}>0$ or $<0$ means that
the dominant contribution comes from the K or K$^{\prime}$ point, respectively,
and that $P_{\sigma}=1$ or $-1$ corresponds to the full valley polarization. 
At $u=0$,
$P_{\uparrow}$ and $P_{\downarrow}$ are of opposite sign and the same magnitude,  
which means no valley polarization.
As $u$ increases,
$P_{\uparrow}$ and $P_{\downarrow}$ become of the same sign, i.e.,
the valley polarization is induced.
Furthermore, the dominant valley can be switched by changing the helicity of CPL;
this is similar to such a property
with resonant weak CPL~\cite{Valley-CPL-Graphene,TMD-ISB1,Valley-CPL-TMD2,Valley-CPL-TMD3}. 
However,
the full valley polarization is not achieved.
The similar results are obtained for the 
other monolayer TMDs driven by CPL
(see Supplemental Note 9 of Supplemental Material~\cite{SM}). 
The absence of the full valley polarization is because 
the energy gaps at the K and K$^{\prime}$ points
are of the order of 1 eV in the nondriven system
and it is difficult to make one of the gaps much smaller than the other.
This result implies that
the valley polarization may be overestimated
in the minimal model~\cite{Floquet-TMD-Dirac1,Floquet-TMD-Dirac2}
which considers only the electronic states near the K and K$^{\prime}$ points.
Note that in that model the full valley polarization is more easily achieved.

Finally, we comment on the experimental realization of our results.
In general, the spin current generated in the SHE can be detected
by the inverse spin Hall effect~\cite{InvSHE1,InvSHE2}.
Furthermore,
the SHE or inverse SHE in a periodically driven system
could be experimentally observed in pump-probe measurements~\cite{Opt-review}.
We should note that
although CPL induces the anomalous Hall effect~\cite{Oka-PRB,Light-AHE-exp},
the spin current generated in the SHE is distinguishable from
the charge current generated in this Hall effect
via the helicity of CPL~\cite{NA-FloquetSHE}.
In our analyses, 
$u=0.1$ corresponds to
$E_{0}\approx 16.1$MVcm$^{-1}$ for MoS$_{2}$ at $\Omega=5$eV,
$19.4$MVcm$^{-1}$ for WS$_{2}$ at $\Omega=6$eV,
and $14.3$MVcm$^{-1}$ for MoTe$_{2}$ or WTe$_{2}$ at $\Omega=5$eV,
where $E_{0}$ is the magnitude of the pump electric field.
In these estimates, we have used
$a_{\textrm{NN}}\approx 3.1$\AA \
for MoS$_{2}$ and WS$_{2}$
and $3.5$\AA \
for MoTe$_{2}$ and WTe$_{2}$~\cite{TMD-TB1}.
From an experimental point of view,
the pump electric field of the order of $10$MVcm$^{-1}$ can be achieved~\cite{Iwai-review}.
As shown in Supplemental Note 4 of Supplemental Material~\cite{SM},
our results remain qualitatively unchanged for smaller $\Omega$. 
Therefore,
we conclude that
our results could be experimentally tested by pump-probe measurements
for the monolayer TMDs driven by CPL.

\section{Acknowledgments}

This work was supported by
JST CREST Grant No. JPMJCR1901, 
JSPS KAKENHI Grant No. JP22K03532, 
and MEXT Q-LEAP Grant No. JP-MXS0118067426.


\begin{references}
\bibitem{TMD1}
  K. F. Mak, C. Lee, J. Hone, J. Shan, and T. F. Heinz,
  Atomically thin MoS$_{2}$: A new direct-gap semiconductor,
  Phys. Rev. Lett. \textbf{105}, 136805 (2010).

\bibitem{TMD2}
  C. Lee, H. Yan, L. E. Brus, T. F. Heinz, J. Hone, and S. Ryu,
  Anomalous lattice vibrations of single- and few-layer MoS$_{2}$,
  ACS Nano \textbf{4}, 2695-2700 (2010).
  
\bibitem{TMD3}
  A. Splendiani, L. Sun, Y. Zhang, T. Li, J. Kim, C.-Y. Chim, G. Galli, and F. Wang,
  Emerging photoluminescence in monolayer MoS$_{2}$,
  Nano Lett. \textbf{10}, 1271-1275 (2010).

\bibitem{TMD-ISB1}
  T. Cao, G. Wang, W. Han, H. Ye, C. Zhu, J. Shi, Q. Niu, P. Tan,
  E. Wang, B. Liu, and J. Feng,
  Valley-selective circular dichroism of monolayer molybdenum disulphide,
  Nat. Commun. \textbf{3}, 887 (2012).

\bibitem{TMD-ISB2}
  D. Xiao, G.-B. Liu, W. Feng, X. Xu, and W. Yao,
  Coupled spin and valley physics in monolayers of MoS$_{2}$
  and other group-VI dichalcogenides,
  Phys. Rev. Lett. \textbf{108}, 196802 (2012). 
  
\bibitem{valleyHall}
  K. F. Mak, K. L. McGill, J. Park, and P. L. McEuen,
  The valley Hall effect in MoS$_{2}$ transistors,
  Science \textbf{344}, 1489 (2014).
  
\bibitem{TopoIns-TMD1}
  X. Qian, J. Liu, L. Fu, and J. Li,
  Quantum spin Hall effect in two-dimensional transition metal dichalcogenides,
  Science \textbf{346}, 1344-1347 (2014).

\bibitem{TopoIns-TMD2}
  S. Wu, V. Fatemi, Q. D. Gibson, K. Watanabe, T. Taniguchi, R. J. Cava,
  and P. J.-Herrero,
  Observation of the quantum spin Hall effect up to 100 kelvin in a monolayer crystal,
  Science \textbf{359}, 76-79 (2018).

\bibitem{Moire-PRL}
  F. Wu, T. Lovorn, E. Tutuc, and A. H. MacDonald,
  Hubbard model physics in transition metal dichalcogenide moir\'{e} bands,
  Phys. Rev. Lett. \textbf{121}, 026402 (2018).

\bibitem{Moire-Nature1}
  E. C. Regan et al.,
  Mott and generalized Wigner crystal states in WSe$_{2}$/WS$_{2}$ moir\'{e} superlattices,
  Nature \textbf{579}, 359 (2020).

\bibitem{Moire-Nature2}
  Y. Tang et al.,
  Simulation of Hubbard model physics in WSe$_{2}$/WS$_{2}$ moir\'{e} superlattices,
  Nature \textbf{579}, 353 (2020).

\bibitem{Moire-Nature3}
  Y. Xu, S. Liu, D. A. Rhodes, K. Watanabe, T. Taniguchi,
  J. Hone, V. Elser, K. F. Mak, and J. Shan,
  Correlated insulating states at fractional fillings of moir\'{e} superlattices,
  Nature \textbf{587}, 214 (2020).

\bibitem{TMD-TB1}
  G.-B. Liu, W.-Y. Shan, Y. Yao, W. Yao, and D. Xiao,
  Three-band tight-binding model for monolayers of
  group-VIB transition metal dichalcogenides,
  Phys. Rev. B \textbf{88}, 085433 (2013).

\bibitem{TMD-TB2}
  S. Bhowal and S. Satpathy,
  Intrinsic orbital and spin Hall effects
  in monolayer transition metal dichalcogenides,
  Phys. Rev. B \textbf{102}, 035409 (2020).

\bibitem{Rashba}
  E. I. Rashba,
  Properties of semiconductors with an extremum loop.
  1. Cyclotron and combinational resonance in a magnetic field perpendicular
  to the plane of the loop,
  Fiz. Tverd. Tela (Leningrad) \textbf{2}, 1224 (1960)
  [Sov. Phys. Solid State \textbf{2}, 1109 (1960)].  
  
\bibitem{TMD-1stPrin1}
  L. F. Mattheiss,
  Band structures of transition-metal-dichalcogenide layer compounds,
  Phys. Rev. B \textbf{8}, 3719 (1973).

\bibitem{TMD-1stPrin2}
  R. Coehoorn, C. Haas, and R. A. de Groot,
  Electronic structure of MoSe$_{2}$, MoS$_{2}$, and WSe$_{2}$.
  II. The nature of the optical band gaps,
  Phys. Rev. B \textbf{35}, 6203 (1987).

\bibitem{TMD-1stPrin3}
  S. Leb\`{e}gue and O. Eriksson
  Electronic structure of two-dimensional crystals from ab initio theory,
  Phys. Rev. B \textbf{79}, 115409 (2009).  
   
\bibitem{Kon-SHE-Pt}
  T. Tanaka, H. Kontani, M. Naito, T. Naito, D. S. Hirashima, K. Yamada, and J. Inoue,
  Intrinsic spin Hall effect and orbital Hall effect in 4$d$ and 5$d$ transition metals,
  Phys. Rev. B \textbf{77}, 165117 (2008).

\bibitem{TMD-SHE-0K}
  L. M. Canonico, T. P. Cysne, A. M.-Sanchez, R. B. Muniz, and T. G. Rappoport,
  Orbital Hall insulating phase in transition metal dichalcogenide monolayers,
  Phys. Rev B \textbf{101}, 161409(R) (2020).   
  
\bibitem{SHE-JETP}
  M. I. D'yakonov and V. I. Perel',
  Possibility of orienting electron spins with current,
  ZhETF Pis. Red. \textbf{13}, 657 (1971)
  [Sov. Phys. JETP Lett. \textbf{13}, 467 (1971)].
  
\bibitem{SHE-Hirsch}
  J. E. Hirsch,
  Spin Hall effect,
  Phys. Rev. Lett. \textbf{83}, 1834 (1999).

\bibitem{Eckstein}
  M. Eckstein and M. Kollar,
  Theory of time-resolved optical spectroscopy
  on correlated electron systems,
  Phys. Rev. B \textbf{78}, 205119 (2008).
  
\bibitem{Tsuji}
  N. Tsuji, T. Oka, and H. Aoki,
  Nonequilibrium steady state of photoexcited correlated electrons
  in the presence of dissipation,
  Phys. Rev. Lett. \textbf{103}, 047403 (2009).

\bibitem{Mikami}
  T. Mikami, S. Kitamura, K. Yasuda, N. Tsuji, T. Oka, and H. Aoki,
  Brillouin-Wigner theory for high-frequency expansion
  in periodically driven systems: Application to Floquet topological insulators,
  Phys. Rev. B \textbf{93}, 144307 (2016).

\bibitem{NA-FloquetSHE}
  N. Arakawa and K. Yonemitsu,
  Symmetry-protected difference between spin Hall and anomalous Hall effects
  of a periodically driven multiorbital metal,
  Commun. Phys. \textbf{6}, 43 (2023).

\bibitem{NA-BCPL}
  N. Arakawa and K. Yonemitsu,
  Light-induced large and tunable valley-selective Hall effect in a centrosymmetric system,
  Phys. Rev. B \textbf{109}, L241201 (2024).

\bibitem{NA-mirror}
  N. Arakawa and K. Yonemitsu,
  Light-induced mirror symmetry breaking and charge transport,
  J. Phys. Soc. Jpn. \textbf{93}, 084701 (2024).
  
\bibitem{HeatBath1}
  M. B\"{u}ttiker,
  Small normal-metal loop coupled to an electron reservoir,
  Phys. Rev. B \textbf{32}, 1846(R) (1985).

\bibitem{HeatBath2}
  M. B\"{u}ttiker,
  Role of quantum coherence in series resistors,
  Phys. Rev. B \textbf{33}, 3020 (1986).

\bibitem{SM}
  See Supplemental Material,
  which includes Refs. \onlinecite{TMD-TB1}, \onlinecite{TMD-TB2},
  \onlinecite{Kon-SHE-Pt}, \onlinecite{Mikami},
  \onlinecite{NA-FloquetSHE},
  \onlinecite{Highw1}, \onlinecite{Highw2}, and 
  \onlinecite{NA-Ru-rot}{--}\onlinecite{Mizo}, for explaining 
  the details of the tight-binding models,
  the Floquet linear-response theory for the SHE,
  the numerical calculations,
  the special models,
  the high-frequency expansion,
  the light-induced group velocities,
  and the light-induced leading-order corrections to $\sigma_{yx}^{\textrm{S}}$,
  and presenting the additional numerical results. 
  
\bibitem{Floquet1}
  J. H. Shirley,
  Solution of the Schr\"{o}dinger equation with a Hamiltonian periodic in time,
  Phys. Rev. \textbf{138}, B979 (1965).

\bibitem{Floquet2}
  H. Sambe,
  Steady states and quasienergies of a quantum-mechanical system
  in an oscillating field,
  Phys. Rev. A \textbf{7}, 2203 (1973).

\bibitem{Kubo}
  R. Kubo,
  Statistical-mechanical theory of irreversible processes. I.
  General theory and simple applications to magnetic and conducting problems,
  J. Phys. Soc. Jpn. \textbf{12}, 570 (1957). 

\bibitem{Highw1}
  A. Eckardt and E. Anisimovas,
  High-frequency approximation for periodically driven quantum systems
  from a Floquet-space perspective,
  New J. Phys. \textbf{17}, 093039 (2015).  
  
\bibitem{Highw2}
  M. Bukov, L. D’Alessio, and A. Polkovnikov,
  Universal high-frequency behavior of periodically driven systems:
  From dynamical stabilization to Floquet engineering,
  Adv. Phys. \textbf{64}, 139 (2015).  
  
\bibitem{Valley-CPL-Graphene}
  W. Yao, D. Xiao, and Q. Niu,
  Valley-dependent optoelectronics from inversion symmetry breaking,
  Phys. Rev. B \textbf{77}, 235406 (2008).
  
\bibitem{Valley-CPL-TMD2}
  H. Zeng, J. Dai, W. Yao, D. Xiao, and X. Cui,
  Valley polarization in MoS$_{2}$ monolayers by optical pumping,
  Nature Nanotech. \textbf{7}, 490 (2012).

\bibitem{Valley-CPL-TMD3}
  K. F. Mak, K. He, J. Shan, and T. F. Heinz,
  Control of valley polarization in monolayer MoS$_{2}$ by optical helicity,
  Nature Nanotech. \textbf{7}, 494 (2012).
  
\bibitem{Floquet-TMD-Dirac1}
  M. Tahir, A. Manchon, and U. Schwingenschl\"{o}gl,
  Photoinduced quantum spin and valley Hall effects,
  and orbital magnetization in monolayer MoS$_{2}$,
  Phys. Rev. B \textbf{90}, 125438 (2014).

\bibitem{Floquet-TMD-Dirac2}
  P. Sengupta and E. Bellotti,
  Photo-modulation of the spin Hall conductivity of
  mono-layer transition metal dichalcogenides,
  Appl. Phys. Lett. \textbf{108}, 211104 (2016).

\bibitem{InvSHE1}
  E. Saitoh, M. Ueda, H. Miyajima, and G. Tatara,
  Conversion of spin current into charge current at room temperature:
  Inverse spin-Hall effect,
  Appl. Phys. Lett. \textbf{88}, 182509 (2006).

\bibitem{InvSHE2}
  S. O. Valenzuela and M. Tinkham,
  Direct electronic measurement of the spin Hall effect,
  Nature \textbf{442}, 176 (2006).

\bibitem{Opt-review}
  A. Kirilyuk, A. V. Kimel, and T. Rasing,
  Ultrafast optical manipulation of magnetic order,
  Rev. Mod. Phys. \textbf{82}, 2731 (2010).  

\bibitem{Oka-PRB}
  T. Oka and H. Aoki,
  Photovoltaic Hall effect in graphene,
  Phys. Rev. B \textbf{79}, 081406(R) (2009).

\bibitem{Light-AHE-exp}
  J. W. McIver et al.,
  Light-induced anomalous Hall effect in graphene,
  Nat. Phys. \textbf{16}, 38 (2020).  
  
\bibitem{Iwai-review}
  Y. Kawakami, H. Itoh, K. Yonemitsu, and S. Iwai,
  Strong light-field effects driven by
  nearly single-cycle 7fs light-field
  in correlated organic conductors,
  J. Phys. B: At. Mol. Opt. Phys. \textbf{51}, 174005 (2018).

\bibitem{NA-Ru-rot}
  N. Arakawa and M. Ogata,
  Origin of the heavy fermion behavior in Ca$_{2-x}$Sr$_{x}$RuO$_{4}$:
  Roles of Coulomb interaction and the rotation of RuO$_{6}$ octahedra,
  Phys. Rev. B \textbf{86}, 125126 (2012).  

\bibitem{Kon-SHE-Ru}
  H. Kontani, T. Tanaka, D. S. Hirashima, K. Yamada, and J. Inoue,
  Giant intrinsic spin and orbital Hall effects in Sr$_{2}M$O$_{4}$ ($M=$Ru, Rh, Mo),
  Phys. Rev. Lett. \textbf{100}, 096601 (2008).

\bibitem{Mizo}
  T. Mizoguchi and N. Arakawa,
  Controlling spin Hall effect
  by using a band anticrossing and nonmagnetic impurity scattering,
  Phys. Rev. B \textbf{93}, 041304(R) (2016).

\end{references}
\end{document}